\documentclass{article}
\usepackage[margin=1in]{geometry}
\usepackage{indentfirst}

\usepackage[utf8]{inputenc} 
\usepackage[T1]{fontenc}    
\usepackage{hyperref}       
\usepackage{url}            
\usepackage{booktabs}       
\usepackage{amsfonts}       
\usepackage{nicefrac}       
\usepackage{microtype}      

\usepackage{bbm}
\usepackage{amsfonts}
\usepackage{amsmath,amsthm}           
\usepackage{mathtools}    
\usepackage{amsopn}
\usepackage{amssymb}
\usepackage{bm}
\usepackage{multirow}
\usepackage{graphicx}
\usepackage{subfigure}
\usepackage{adjustbox}
\usepackage{xcolor}
\usepackage[vlined,linesnumbered,ruled]{algorithm2e}
\usepackage{tabularx}
\usepackage{ragged2e}
\usepackage{booktabs}
\usepackage{enumitem}
\usepackage{biblatex}
\usepackage{multirow}
\usepackage{multicol}
\usepackage{tabularx}
\usepackage[normalem]{ulem}
\usepackage[linesnumbered,ruled,vlined]{algorithm2e}
\usepackage{arydshln}
\usepackage{array}
\newcommand\pref[1]{(\ref{#1})}

\allowdisplaybreaks

\newcolumntype{L}{>{\RaggedRight\hangafter=1\hangindent=0em}X}

\allowdisplaybreaks

\title{DimeRec: A Unified Framework for Enhanced Sequential Recommendation via Generative Diffusion Models}

\author{
Wuchao Li\\
University of Science and Technology of China \\
\texttt{liwuchao@mail.ustc.edu.cn}
\and
Rui Huang\\
Kuaishou Inc.\\
\texttt{huangrui06@kuaishou.com}
\and 
Haijun Zhao\\
Sun Yat-Sen University \\
\texttt{zhaohj23@mail2.sysu.edu.cn}
\and
Chi Liu\\
Kuaishou Inc.\\
\texttt{liuchi05@kuaishou.com}
\and 
Kai Zheng\\
Kuaishou Inc. \\
\texttt{zhengkai@kuaishou.com}
\and
Qi Liu\\
University of Science and Technology of China \\
\texttt{qiliu67@mail.ustc.edu.cn}
\and
Na Mou\\
Kuaishou Inc. \\
\texttt{mouna@kuaishou.com}
\and
Guorui Zhou\\
Kuaishou Inc. \\
\texttt{zhouguorui@kuaishou.com}
\and
Defu Lian\\
University of Science and Technology of China \\
\texttt{liandefu@ustc.edu.cn}
\and
Yang Song\\
Independent \\
\texttt{ys@sonyis.me}
\and
Wentian Bao\\
Independent \\
\texttt{wb2328@columbia.edu}
\and
Enyun Yu\\
Independent \\
\texttt{yuenyun@126.com}
\and
Wenwu Ou\\
Independent \\
\texttt{ouwenwu@gmail.com}
\and
}
\date{}

\begin{document}


\maketitle
\begin{abstract}
Sequential Recommendation (SR) plays a pivotal role in recommender systems by tailoring recommendations to user preferences based on their non-stationary historical interactions. Achieving high-quality performance in SR requires attention to both item representation and diversity. However, designing an SR method that simultaneously optimizes these merits remains a long-standing challenge. In this study, we address this issue by integrating recent generative Diffusion Models (DM) into SR. DM has demonstrated utility in representation learning and diverse image generation \cite{chen2024deconstructing}. Nevertheless, a straightforward combination of SR and DM leads to sub-optimal performance due to discrepancies in learning objectives (recommendation vs. noise reconstruction) and the respective learning spaces (non-stationary vs. stationary). To overcome this, we propose a novel framework called DimeRec (\textbf{Di}ffusion with \textbf{m}ulti-interest \textbf{e}nhanced \textbf{Rec}ommender). DimeRec synergistically combines a guidance extraction module (GEM) and a generative diffusion aggregation module (DAM). The GEM extracts crucial stationary guidance signals from the user’s non-stationary interaction history, while the DAM employs a generative diffusion process conditioned on GEM’s outputs to reconstruct and generate consistent recommendations. Our numerical experiments demonstrate that DimeRec significantly outperforms established baseline methods across three publicly available datasets. Furthermore, we have successfully deployed DimeRec on a large-scale short video recommendation platform, serving hundreds of millions of users. Live A/B testing confirms that our method improves both users' time spent and result diversification.

\end{abstract}



\section{Introduction}
\label{sec:introduction}

Sequential Recommendation (SR) is a fundamental component of modern industrial recommender systems, providing accurate and diversified candidate items for subsequent stages. While mainstream SR solutions focus on analyzing historical interactions between users and the recommendation system (RS), their primary goal is to predict the next item a user is likely to prefer. 
Recent scholarly efforts, exemplified by works such as GRU4Rec\cite{hidasi2015session}, BERT4Rec\cite{Sun_Liu_Wu_Pei_Lin_Ou_Jiang_2019}, and SASRec\cite{Kang_McAuley_2018}, have harnessed powerful sequence models from the field of Natural Language Processing (NLP), including Recurrent Neural Networks (RNNs) and Transformers. Relying solely on a fixed user embedding may limit the expressive power when retrieving diverse items. While multi-interest approaches like MIND \cite{Li_Liu_Wu_Xu_Zhao_Huang_Kang_Chen_Li_Lee_2019} and ComiRec \cite{Cen_Zhang_Zou_Zhou_Yang_Tang_2020} learn multi-representation for users, they are also based on fixed dense vectors for representation, which still belongs to discriminative paradigm \cite{yang2023generate}. This limitation restricts their ability to adequately represent diversity and uncertainty, failing to grasp the underlying distribution of user interests accurately. Another significant drawback of deterministic SR methods is their susceptibility to exposure bias. This issue arises because these methods presuppose that items with which users have interacted most are the most relevant, thereby placing items with fewer interactions at a disadvantage. 

To mitigate these limitations, there has been a pivot towards the adoption of robust generative models like VAEs and GANs. These models are proficient at capturing the distribution of item representations, effectively reducing exposure bias \cite{Zhao_Zhao_Zhao_Liu_Sheng_Zhou_2021, Xie_Liu_Zhang_Lu_Wang_Ding_2021, Ren_Liu_Li_Zhao_Wang_Ding_Wen_2020}. Nevertheless, the application of these models in SR is constrained by the information bottleneck characteristic of VAEs on item representations \cite{Alemi_Fischer_Dillon_Murphy_2016} and the inherent instability of GANs\cite{Salimans_Goodfellow_Zaremba_Cheung_Radford_Chen}. Diffusion Models (DMs)\cite{Ho_Jain_Abbeel_2020, Dhariwal_Nichol_2021}, a novel paradigm of generative models, have garnered substantial acclaim in recent research within the vision and NLP fields due to their impressive successes. A significant advantage of DMs is the tractable posterior within the generation process, facilitating flexible noise schedule adjustment. This adaptability enhances the model's representational capabilities \cite{chen2024deconstructing} and stabilizes the training process, thereby positioning DMs as potentially invaluable assets for devising advanced SR techniques to enhance item representation and diversified results.

Initially conceptualized for noise injection and denoising operations within continuous spaces, DMs face difficulty addressing the discrete spaces inherent to SR, since there are enormous meaningless item IDs and candidate items also vary with time. Recently, researchers in recommender systems \cite{Li_Sun_Li_2023, yang2023generate} have introduced diffusion into the item latent space to address this issue. However, these methods may not be optimal, preventing these models from fully utilizing diffusion. The paradigm of these models is aimed at generating the embedding of the next item, using the user's entire historical behavior sequence as the condition for denoising, and fitting the distribution across the entire item space. Considering the sparsity of user and item interactions in real-world scenarios, this is undoubtedly very challenging. This difficulty also necessitates setting a large maximum number of denoising steps to accurately generate representations in the complex item representation space. Another drawback of existing models is the lack of a comprehensive understanding of the principles of diffusion models, leading to improper integration when applied to recommendation systems. Specifically, during the loss training process, the optimization directions of the classification loss for the recommendation model and the reconstruction loss for the diffusion model are inconsistent, resulting in suboptimal recommendation performance. Furthermore, these methods involve models trained in an end-to-end manner, where the initial representations of items are disorganized, implying that their distribution may be also meaningless. Training diffusion models directly in such a space without any assistance is difficult.

To address the aforementioned challenges, this paper introduces DimeRec (Diffusion with Multi-interest Enhanced Recommender), a novel approach that shifts the granularity of recommendation generation from individual items to the user’s next area of interest. Notably, instead of relying on behavioral sequences as the denoising condition, we leverage the user’s multiple interests. These interests serve as high-level abstractions of the vast array of items, and the interest space tends to be more stable compared to the item space, similar to the fixed pixel values in the original application of diffusion in images. This stability facilitates learning for diffusion. To address the issue of simultaneously optimizing the recommendation loss and the diffusion model's loss, we introduce a new noise space where both types of loss can be optimized concurrently. Furthermore, we introduce an independent guidance loss alongside the diffusion module. This additional loss function enhances the stability and effectiveness of learning item representations and the distribution acquired through the diffusion process. Empirical evaluations conducted on public datasets and live A/B tests within industrial recommendation systems demonstrate that DimeRec effectively enhances the performance of sequence-based recommendations. The contributions can be summarized as:

\begin{itemize}[leftmargin=*]
    \item We propose a unified framework for enhanced sequence recommendation via generative diffusion models, named \textbf{DimeRec}, which transforms the traditional approach of generating the next item into generating the next user interest. Stationary interest guidance instead of non-stationary interaction history helps DM for better reconstruction and representation learning. 
    \item We propose a new noise space, along with a novel loss function, to make the application of diffusion models in recommendation systems more effective. The loss function includes a guiding loss, which enhances the stability of the representation learning after introducing the diffusion model.
    \item We conduct extensive experiments and verify the improvement of \textbf{DimeRec} compared to multiple baseline methods on three public datasets. Moreover, online A/B experiments show that \textbf{DimeRec} can significantly improve users' time spent and satisfaction of users, as well as the recommendation diversity.
\end{itemize}

\section{Related Work}
\label{sec:releated_work}

\subsection{Sequential Recommendation}
Classic SR methods such as FPMC \cite{Rendle_Freudenthaler_Schmidt-Thieme_2010} and Fossil \cite{Ruining_Julian_2016} utilize Markov Chains to model dynamic transitions within interaction sequences. With the advancement of deep learning, recent SR methods have employed more powerful deep neural networks to capture the cross-dependencies between items in sequence. For instance, GRU4Rec \cite{hidasi2015session} utilizes the gated recurrent unit (GRU) to capture temporal dependencies between sequences. Caser \cite{Tang_Wang_2018} employs a convolutional neural network (CNN) and incorporates both horizontal and vertical convolutional filters to capture union-level and point-level patterns in the sequences, respectively. SASRec \cite{Kang_McAuley_2018} turns to a more efficient multi-layer transformer to model sequence interactions. BERT4Rec \cite{Sun_Liu_Wu_Pei_Lin_Ou_Jiang_2019} further extends the model structure based on bidirectional transformer layers and leverages the Cloze \cite{Devlin_Chang_Lee_Toutanova_2019} task to encode behavior sequences. These methods have demonstrated effective and efficient modeling of patterns within behavioral sequences, providing guidance for the sequence modeling in our work.

\subsection{Diffusion Models for Recommendation}
Recently, DMs have been integrated into RS research, DiffRec \cite{Du_Yuan_Huang_Zhao_Zhou_2023} incorporates an additional transformation to map items from discrete space to continuous space. Notably, the noise and denoising methods are exclusively applied to the target item. DiffuRec \cite{Li_Sun_Li_2023} utilizes truncated scheduled Gaussian noise in the forward process to corrupt the target item embedding. Additionally, it introduces a Transformer-based Approximator structure to predict $x_0$ in the reverse process, allowing iterative reconstruction of the target item embedding. Furthermore, \citeauthor{Wang_Xu_Feng_Lin_He_Chua_2023} expand the standard diffusion models to the latent space and introduce a time-sensitive approach to model the probability of the user's interaction sequence. Recently, DreamRec\cite{yang2023generate} abandons negative sampling and completely transforms the recommendation task into a learning-to-generate paradigm. Although DreamRec has achieved comparable results on several specific datasets, it is challenging to apply it in large-scale scenarios, especially in industrial settings, because it lacks the generalization effect brought by negative sampling. DCDR\cite{lin2024discrete} designs a new discrete conditional diffusion re-ranking framework, which progressively generates and optimizes recommendation lists through a diffusion process in discrete space. However, it is only applicable to recommendation re-ranking and not to the retrieval stage, which is the focus of this paper.
\section{Preliminary}
\label{sec:preliminary}





\textit{Diffusion Models} consist of the forward process and the reverse process.

In the forward process, DMs gradually add Gaussian noise to a sample $x_0 \sim q(x)$ with a maximum step of $T$, each step $t \in [1, T]$ subjects to the following transition
\begin{equation}
    q(x_t | x_{t-1}) = \mathcal{N}(x_t ; \sqrt{1 - \beta_t} x_{t-1}, \beta_t \mathbb I)
\end{equation}
where $\beta_t$ follows the pre-designed variance schedule, $\mathbb{I}$ denotes the identity matrix. This process can be efficient as a multi-step transition can be tractable as
\begin{align}
    q(x_t | x_0) &= \mathcal{N}(x_t; \sqrt{\bar{\alpha}_t} x_0 , (1-\bar{\alpha}_t) \mathbb{I}) \\
    \text{where} \ \bar{\alpha}_t &= \prod_{i=1}^t (1 - \beta_i)
\end{align}
Utilizing the reparameterization trick \cite{Kingma_Salimans_Welling_2015}, we have
\begin{equation}
    \label{eq:eq1}
    x_t = \sqrt{\bar{\alpha}_t} x_0 + \sqrt{1 - \bar{\alpha}_t} \epsilon
\end{equation}

In the reverse process, DMs learn a parameterized joint distribution $p_{\theta}(x_{0:T})$ defined by a Markov chain as
\begin{align}
    p_{\theta}(x_{0:T}) &= p(x_{T}) \prod_{t=1}^{T}p_{\theta}(x_{t-1} | x_t) \\
    p_{\theta}(x_{t-1} | x_t) &= \mathcal{N}(x_{t-1}; \mu_{\theta}(x_t, t), \sigma_{\theta}(x_t, t))
\end{align}

Here, $p(x_T) \sim \mathcal{N}(x_T; \mathbf{0}, \mathbb{I})$ is the starting point drawn from standard Gaussian distribution, $\mu_{\theta}(x_t, t)$ and $\sigma_{\theta}(x_t, t)$ parameterize the mean and variance at step $t$, respectively. Thanks to a condition on $x_0$, the reverse conditional probability can be tractable using the Bayes rule:
\begin{align}
    p_{\theta}(x_{t-1} | x_t, x_0) &= \mathcal{N}(x_{t-1}; \tilde{\mu}_t(x_t, x_0), \tilde{\beta}_t \mathbb{I}) \\
    \label{eq:eq2}\tilde{\mu}_t(x_t, x_0) &= \frac{\sqrt{1 - \beta_t} (1 - \bar{\alpha}_{t-1})}{1 - \bar{\alpha}_t} x_t + \frac{\sqrt{\bar{\alpha}_{t-1}} \beta_t}{1 - \bar{\alpha}_t} x_0 \\
    \tilde{\beta}_t &= \frac{1 - \bar{\alpha}_{t-1}}{ 1- \bar{\alpha}_{t}} \beta_t
\end{align}

Following the formulation of DDPM  \cite{Ho_Jain_Abbeel_2020}, the variance is set untrained $\tilde{\beta}_t = \sigma_t^2 \mathbb{I}$. However, the term $x_0$ is the groud-truth which is not available in the reverse phase, we can substitute $x_0$ with $x_t$ via Eq. \pref{eq:eq2} by combining Eq. \pref{eq:eq1}:
\begin{align}
    \tilde{\mu}_t(x_t, x_0) &= \frac{\sqrt{1 - \beta_t} (1 - \bar{\alpha}_{t-1})}{1 - \bar{\alpha}_t} x_t + \frac{\sqrt{\bar{\alpha}_{t-1}} \beta_t}{1 - \bar{\alpha}_t} {\frac{1}{\sqrt{\bar{\alpha}_t}} \left(x_t - \sqrt{1 - \bar{\alpha}_t} \epsilon_t \right)} \\
    &= \frac{1}{ \sqrt{1 - \beta_t}} \left( x_t - \frac{\beta_t}{\sqrt{1 - \bar{\alpha}_t}} \epsilon_t \right)
\end{align}
Thus, a denoising model $f_{\theta}(x_t, t)$ can be trained to predict the noise term $\epsilon_t$ as $x_t$ is available at timestep $t$.

Efficient training of DMs is possible by optimizing the simplified MSE loss instead of the original variational lower bound (VLB) as
\begin{equation}
    \mathcal{L}_{\text{simple}} =  \mathbb E_{t, x_0, \epsilon} \left[ \lVert \epsilon - f_{\theta}(x_t, t) \rVert_2^2\right]
\end{equation}


\section{Methodology}
\label{sec:method}

\subsection{Framework Overview}
\label{sec:framework_overview}

\begin{figure*}[h]
  \centering
  \includegraphics[width=0.92\textwidth]{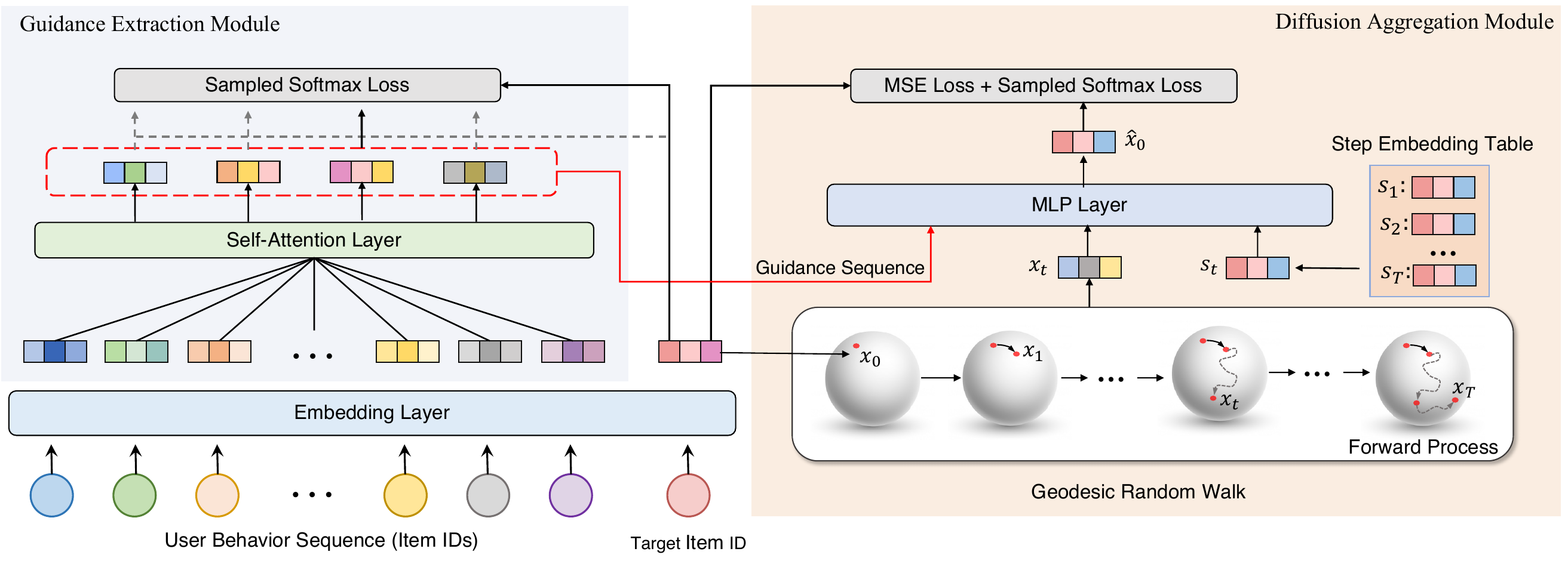}
  \caption{Architecture of model-based DimeRec. In the left part, we use the Self-Attention backbone for guidance extraction from the raw behavior sequence. The right part shows the DAM structure, we employ Geodesic Random Walk on the sphere to add noise to the embedding of the target item. Under the guidance sequence of multi-interest, we restore the embedding to its original state using a Multi-Layer Perceptron (MLP). The two parts are trained jointly in a multi-task way.}
  \label{fig:arch}
\end{figure*}

The overall framework consists of two inter-dependent modules: 1) the guidance extraction module $\mathbf{GEM}_{\phi}(\cdot)$ and 2) the diffusion aggregation module $\mathbf{DAM}_{\theta}(\cdot)$, where $\phi$ and $\theta$ are the corresponding parameters. 

\textbf{In the first stage}, the guidance extraction module (GEM) takes in the user's original behavior sequence $\mathcal{S}^u$ and extracts out an abstract user embedding sequence $\mathbf{g}^u$: 
\begin{equation}
    \mathbf{g}^u = \mathbf{GEM}_{\phi}(\mathcal{S}^u).
\end{equation}

Existing diffusion-based models directly extract and encode information from the user's entire behavior sequence and then use it as guidance for the diffusion module's denoising process. However, compared to the user interests, the user's behavior history is non-stationary, and its vague information used directly to guide the diffusion model denoising can impair the final recommendation effect. Therefore, we propose for the first time to use a multi-interest model \cite{Cen_Zhang_Zou_Zhou_Yang_Tang_2020, Li_Liu_Wu_Xu_Zhao_Huang_Kang_Chen_Li_Lee_2019} as our GEM, which can extract the user's relatively stationary interests through rules designed for specific tasks, powerful dynamic routing \cite{Sabour_Frosst_Hinton_2017} or self-attentive \cite{Lin_Feng_Nogueira_Santos_Yu_Xiang_Zhou_Bengio_Watson} mechanisms, providing clearer information during the reversing process.

Moreover, existing models only train on the diffusion module, leading to unstable representation learning of items and difficulty in converging to an optimal effect. This is because the guidance extracted by the GEM does not fit well with the user due to a lack of training. Therefore, we propose a guidance loss that is unrelated to diffusion module and solely trains the GEM module, which can make the learning of item representation more stable and the extracted guidance more precise.

\textbf{In the second stage}, the diffusion aggregation module (DAM) can make full use of diffusion models \cite{Ho_Jain_Abbeel_2020} to aggregate the aforementioned user embedding sequence $\mathbf{g}^u$ in latent embedding space to generate the final user embedding $\mathbf{e}_u$:
\begin{equation}
    \mathbf{e}_u = \mathbf{DAM}_{\theta}(\mathbf{g}^u, x^t, t)
\end{equation}
where $x_t$ is the noised embedding of the target item, and $t$ is used to obtain the step embedding. 

An independent GEM module can actually be used for recommendations, which is the approach adopted by most industrial scenarios. However, these models all belong to the classification paradigm, which can easily lead to recommendations lacking diversity. In contrast, the generative paradigm used by diffusion models learn a broader distribution of user interests, which can bring more diverse items to users. Furthermore, recent research\cite{chen2024deconstructing} has shown that diffusion also plays a certain role in enhancing representation learning, inspiring us that it can match user interests more accurately while bringing diversity. However, directly combining the two modules will lead to incompatibility between the classification paradigm's loss and the generative paradigm's loss. Existing models either discard the former\cite{yang2023generate} or the latter\cite{Li_Sun_Li_2023}, which undoubtedly undermines the model's capabilities. We address this by constraining the item representation space to a sphere and, following previous work\cite{de2022riemannian}, introduce Geodesic Random Walk, allowing these two types of loss to be optimized simultaneously.

\subsection{Guidance Extraction Module}
\label{sec:general_guidance_module}

\textbf{Rule-based GEM.} There is no learnable parameter in $\textbf{GEM}_{\phi}(\cdot)$ in this case. In the real scene, we can design rich rules based on actual business, for example, we can keep the most popular items based on exposure, empirical XTRs, and other statistics or we can explicitly balance the exploration and exploitation (EE) by taking in some promising but may be underestimated items. Rule-based GEM is straightforward, but in real-world scenarios, they can provide the most effective guidance to help build a controllable sequential recommendation directly under human knowledge. In this paper, a basic solution is to cut down $\mathcal{S}^u$ from size $N$ to a smaller size $K$. We can obtain the guidance by performing list slicing on $\mathcal{S}^u$ and encoding to continuous space as:
\begin{equation}
    \mathbf{g}^u = \mathcal{F}(\mathcal{S}^u[-K:])
\end{equation}
which extracts the latest $K$-sub sequence $\mathbf{g}^u \in \mathbb{R}^{K \times d}$, where $\mathcal{F}$ denotes the encoding function, and $d$ is the size of embedding.

\textbf{Model-based GEM}, following the prior works \cite{Lin_Feng_Nogueira_Santos_Yu_Xiang_Zhou_Bengio_Watson,Cen_Zhang_Zou_Zhou_Yang_Tang_2020}, we apply the self-attentive mechanism to our guidance extraction module. The original user behavior sequence $\mathcal{S}^u$ is first encoded by $\mathcal{F}$ as $\mathcal{H}^u = \mathcal{F}(\mathcal{S}^u)$. The self-attentive method extracts the attention weights $\mathbf{A} \in \mathbb{R}^{N \times K}$ using a two-layer \textbf{MLP} with \textbf{Tanh} hidden activation as
\begin{equation}
    \mathbf{A} = \text{Softmax}\left(\textbf{MLP}([4d, K], \textbf{Tanh})(\mathcal{H}^u + \mathcal{P})  \right)
\end{equation}
where $K$ is the guidance length and much smaller than $N$ to make the following DAM module efficient, and $\mathcal{P}$ denotes the standard global positional embeddings \cite{Vaswani_Shazeer_Parmar_Uszkoreit_Jones_Gomez_Kaiser_Polosukhin_2017} which captures the order of sequence. The final guidance sequence $\mathbf{g}^u \in \mathbb{R}^{K \times d}$ is obtained by performing a weighted sum on $\mathcal{H}^u$ as follows:
\begin{equation}
    \mathbf{g}^u = \mathbf{A}^{T} \mathcal{H}^u
\end{equation}

\textbf{Loss Function.} Denote the learnable parameters in MLP as $\phi$, we can optimize the GEM model based on the matching level between the guidance sequence $\mathbf{g}^u$ and target item embedding $e_{a}$ by introducing an $\arg \max$ operator, which selects the most similar guidance embedding in embedding space:

\begin{equation}
    g_u = \mathbf{g}^u[:, \arg \max (\mathbf{g}^u \mathbf{e}_a)]
\end{equation}

The GEM module can be trained with the following objective function
\begin{equation}
    \mathcal{L}_{gem}(\phi) = \frac{1}{|\mathcal{U}|} \sum_{a \in \mathcal{I}_u} - \log \left(\frac{\exp(g_u \cdot \mathbf{e}_{a})}{\exp(g_u \cdot \mathbf{e}_{a}) + \sum_{i^{-} \in \mathcal{I}_{\text{sample}} \exp(g_u \cdot \mathbf{e}_{i^{-}})}} \right)
\end{equation}
Which employs sampled softmax loss technique following the common retrieval practices in SR.

\subsection{Diffusion Aggregation Module}
\label{sec:generative_aggregation_module}

DAM extends vanilla diffusion models (DMs) to learn guidance controllable sequential recommendation in the latent item embedding space.

\textbf{Denoising Model with Guidance.} Given the prior guidance sequence $\mathbf{g}^u$, the diffusion models can be implemented with a conditional denoising model $f_{\theta}(x_t, t, \mathbf{g}^u)$ to predict the target $x_0$ instead of the noise. We model the conditional denoising model $f_{\theta}(x_t, t, \mathbf{g}^u)$ with a simple Multilayer Perceptron:
\begin{equation}
    \hat{x}_0 = f_{\theta}(x_t, t, \mathbf{g}^u) = MLP(x_t, t, \mathbf{g}^u)
\end{equation}

\textbf{Loss Function.} As we predict $x_0$, the simplified training objective learns the following MSE reconstruction objective:

\begin{equation}
    \mathcal{L}_{recon} (\theta) = \mathbb E_{t, x_0, \epsilon} \left[ \lVert x_0 - f_{\theta}(x_t, t, \mathbf{g}^u) \rVert_2^2\right]
\end{equation}

For the retrieval task, it's straightforward to optimize the sampled softmax loss \cite{Jean_Cho_Memisevic_Bengio_2014} on $\hat x_0$ as

\begin{equation}
    \mathcal{L}_{ssm} (\theta) = \frac{1}{|\mathcal{U}|} \sum_{a \in \mathcal{I}_u} - \log \left(\frac{\exp(\hat x_0 \cdot \mathbf{e}_{a})}{\exp(\hat x_0 \cdot \mathbf{e}_{a}) + \sum_{i^{-} \in \mathcal{I}_{\text{sample}} \exp(\hat x_0 \cdot \mathbf{e}_{i^{-}})}} \right)
\end{equation}

Combining all together, the \textbf{DimeRec} framework learns the GEM and DAM jointly by optimizing the following loss:
\begin{equation}
    \label{eq:total_loss}
    \mathcal{L} = \mathcal{L}_{\text{gem}} + \lambda \mathcal{L}_{\text{recon}} + \mu \mathcal{L}_{\text{ssm}}
\end{equation}
Where $\lambda$ and $\mu$ are coefficients that balance the diffusion reconstruction loss and diffusion sampled softmax loss. 

\begin{figure}[htb]
	\centering
 	\includegraphics[width=0.22\textwidth]{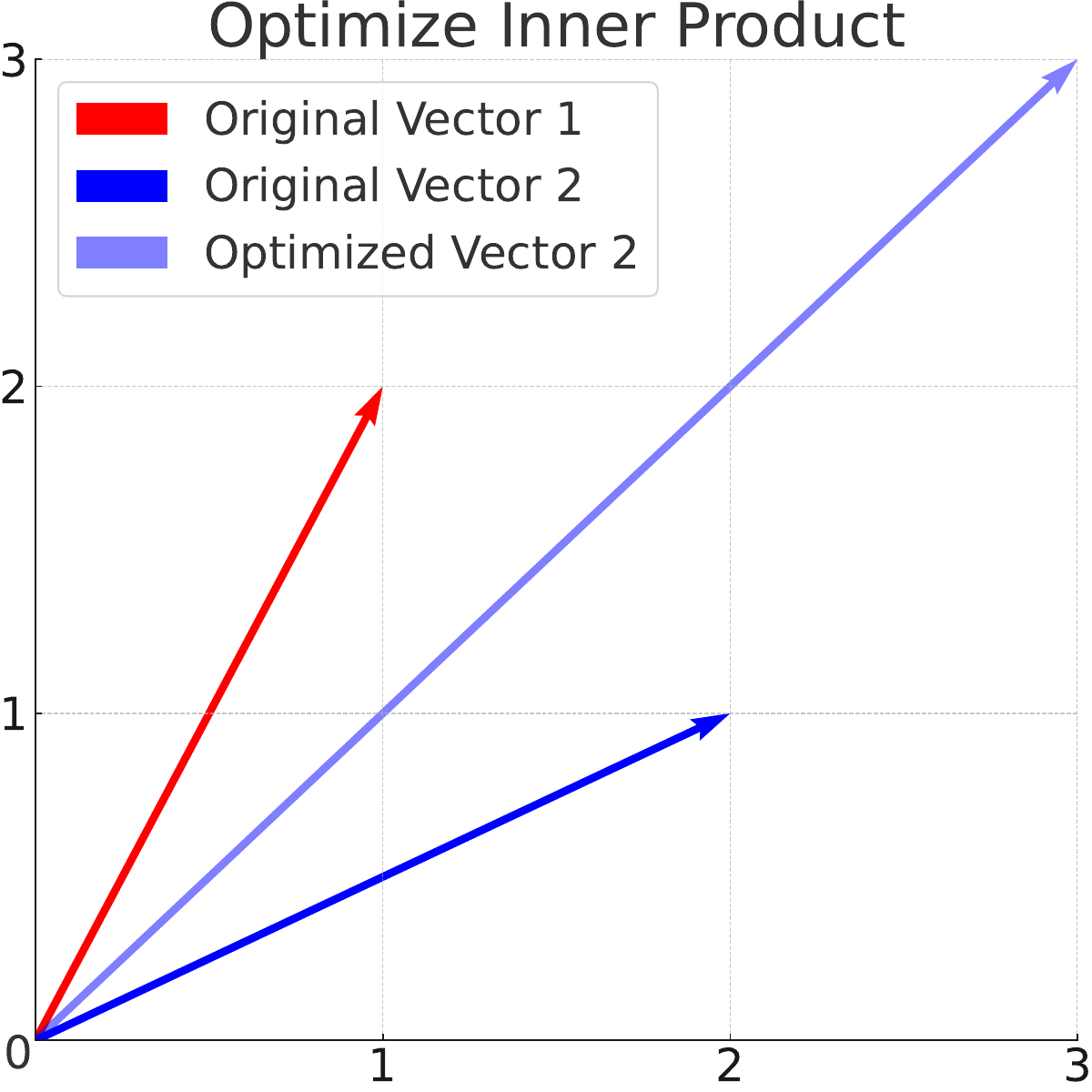}
 	\includegraphics[width=0.22\textwidth]{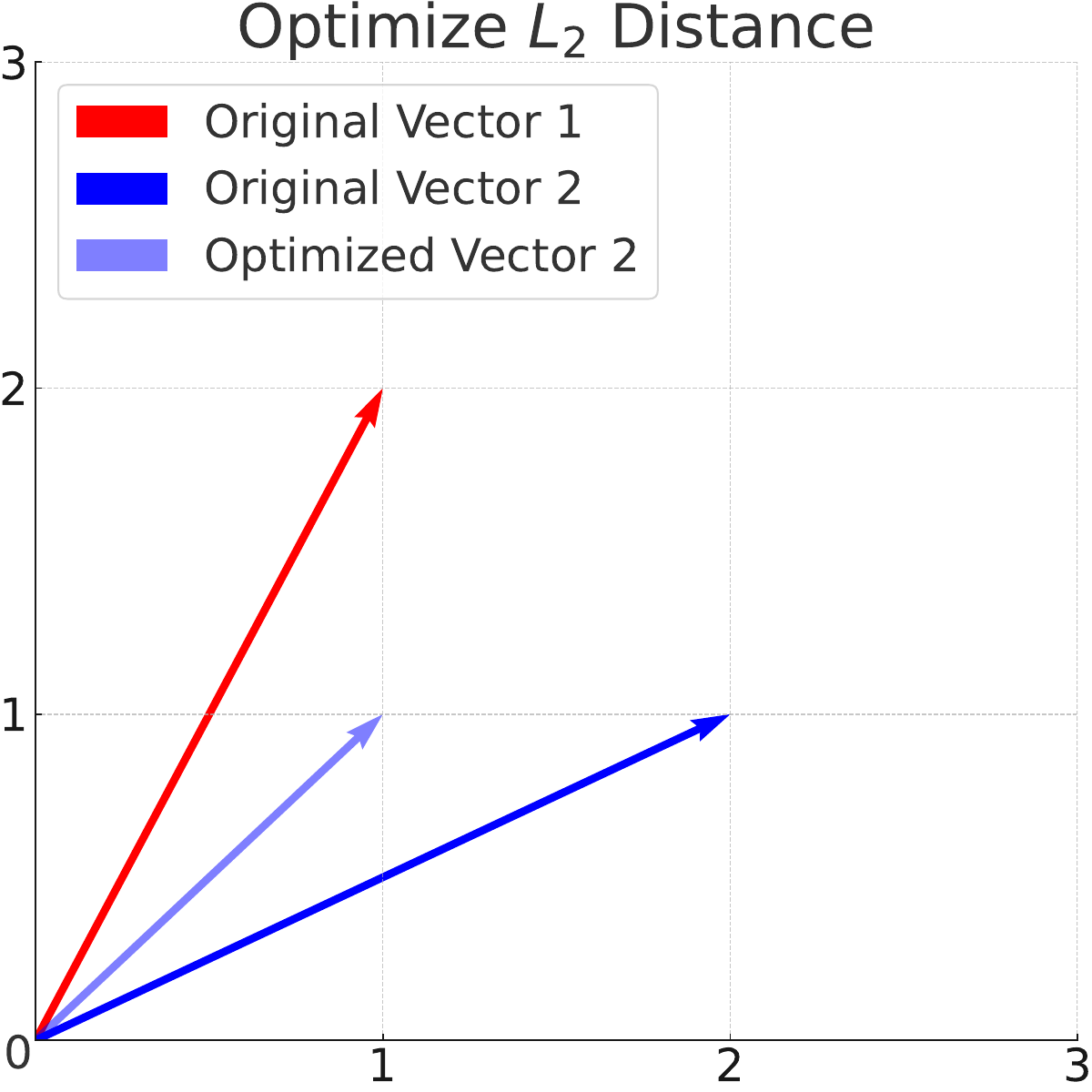}
	\caption{Example of Optimization Inconsistency.}
	\label{fig:optimize}
\end{figure}

\textbf{Geodesic random walk.}
Existing methods add noise to item embeddings in Euclidean space, however, this can lead to a divergence in the optimization directions between the reconstruction loss and the recommendation loss. 
This is mainly because optimizing to minimize the $L_2$ distance usually involves reducing the magnitude of the vectors, while optimizing to increase the inner product typically consists of increasing the magnitude of the vectors. This makes it very challenging to optimize both simultaneously, as illustrated in Figure \ref{fig:optimize}. In the left figure, the optimization goal is to increase the inner product of vector 2 with vector 1. After optimization, the inner product increases, but the $L_2$ distance also increases. In the right figure, the situation is completely opposite, with both the $L_2$ distance and the inner product decreasing. Training them simultaneously can lead to incompatibility in optimization, which we have also verified in our ablation study. An obvious solution is to map the item representations onto a spherical space, where their $L_2$ norm becomes a constant. But how can we add noise to the representations in the spherical space? Recent research on diffusion in Riemannian spaces \cite{de2022riemannian} tells us that we can adopt the Geodesic Random Walk. We provide the pseudocode for the overall training procedure Algorithm \ref{alg:algo_2} in appendix. According to the central limit theorem for manifolds, for a spherical manifold, if the steps of the geodesic random walk are small and isotropically distributed (i.e., uniformly distributed in all directions), the distribution of the endpoints of such walks can approximate Gaussian noise in the tangent space at the mean point of the walks \cite{hsu2002stochastic}. This inspires us that we can directly utilize the conclusions of DDPM\cite{Ho_Jain_Abbeel_2020} for the reverse process without losing effectiveness. The only thing we need to do is to map the generated vectors onto the spherical space.
\textbf{DimeRec} first applies the GEM to extract the guidance sequence. 

\subsection{Serving Strategy}
\label{sec:serving_strategy}
In DAM, we sample an initial user embedding from standard Gaussian distribution, the guidance sequence and user embedding are then fed into the iterative reverse process to obtain the final user embedding $\mathbf{e}_u$, which we summarize the pseudocode Algorithm \ref{alg:algo_3} in appendix. Finally, it performs an independent ANN search on the global item pool to get the Top-$N$ candidates.



\subsection{Complexity Analysis}
The majority of the model's parameters come from item embeddings, with a complexity of $O(Nd)$, where $N$ is the number of items and $d$ is the dimension of the embeddings. For the time complexity of inference, the complexity of the forward network is $O(nd^2)$, and the complexity of self-attention is $O(n^2d)$, where $n$ represents the length of the user sequence, consistent with the original SASRec and DiffuRec papers. For diffusion-based models, we analyze the scenario where the number of steps is set to 1. In offline experimental scenarios, as the number of reverse steps increases, the inference time also increases linearly. However, in online scenarios, the time for full-link inference for each query does not exceed 1ms, and the time taken for DimeRec inference is only a small portion, so even increasing the number of steps has little impact on efficiency.

\section{Experiments}
\label{sec:experiment}
We conduct extensive experiments on both offline real-world public datasets and online platform to answer the following questions:
\begin{itemize}
\item[\textbf{RQ1}] What is the overall performance of \textbf{DimeRec} compared to other popular methods?
\item[\textbf{RQ2}] Why do we introduce Diffusion into recommendation systems?
\item[\textbf{RQ3}] What is the effectiveness of each component in \textbf{DimeRec}?
\item[\textbf{RQ4}] How to set the hyper-parameters in loss function \ref{eq:total_loss} for \textbf{DimeRec}?
\item[\textbf{RQ5}] How does our proposed \textbf{DimeRec} perform in the online industrial scenario?
\end{itemize}

\subsection{Experiment Settings}
\subsubsection{Dataset} We select three widely-used real-world public datasets to evaluate the effectiveness of DimeRec:
\textbf{YooChoose}\footnote{https://recsys.acm.org/recsys15/challenge/}, collected from the YooChoose e-commerce platform. Each training sample is truncated at length 10.
\textbf{KuaiRec}\footnote{https://cseweb.ucsd.edu/~jmcauley/datasets/amazon/links.html}, collected from a leading short video sharing platform. Each training sample is truncated at length 10.
\textbf{ML-10M}\footnote{http://files.grouplens.org/datasets/movielens/}, collected from the movie website by GroupLens Research. Each training sample is truncated at length 70.


\begin{table*}[t]
\centering
\caption{Comparison with Baselines w.r.t HR@10, NDCG@10, HR@20, NDCG@20, HR@50 and NDCG@50. All experiments are conducted five times, and the table shows the mean and standard deviation of the results. The best results are highlighted in bold. * indicates that the model has shown significant improvement relative to the second-best model (t-test P<.05).} 
\renewcommand{\arraystretch}{1}
\newcolumntype{Y}{>{\centering\arraybackslash}X}
\begin{tabularx}{\linewidth}{Y|Y|YY|YY|YY} 
\toprule
Dataset & Method                                                   & HR@10 (\%)       & NDCG@10(\%)  & HR@20(\%)                                            & NDCG@20(\%)        & HR@50(\%)  & NDCG@50(\%)                                      \\ 
\midrule
\multirow{8}{*}{YooChoose}
& SASRec                  & 2.73$\pm$0.05 	& 1.38$\pm$0.04	& 3.70$\pm$0.06 &	1.63$\pm$0.10	& 5.18$\pm$0.08	& 1.92$\pm$0.03 \\
& GRU4Rec                 &2.97$\pm$0.11 	& 1.56$\pm$0.03	& 4.18$\pm$0.05 &	1.86$\pm$0.05	& 6.32$\pm$0.13	& 2.29$\pm$0.04 \\
& MIND                 & 3.71$\pm$0.09 	& \textbf{1.98$\pm$0.02}	& 4.87$\pm$0.12 &	2.27$\pm$0.07	& 6.55$\pm$0.15	& 2.60$\pm$0.13 \\
& ComiRec                  & 3.53$\pm$0.07 	& 1.39$\pm$0.09	& 5.40$\pm$0.11 &	1.86$\pm$0.05	& 8.75$\pm$0.12	& 2.52$\pm$0.09 \\
& Mult-VAE             & 3.58$\pm$0.08 	& 1.76$\pm$0.07	& 4.39$\pm$0.07 &	2.16$\pm$0.12	& 6.98$\pm$0.20	& 2.86$\pm$0.12 \\
& DreamRec                 & 1.79$\pm$0.02 	& 0.83$\pm$0.03	& 3.45$\pm$0.07 &	1.24$\pm$0.02	& 6.94$\pm$0.07	& 1.92$\pm$0.07 \\
& DiffuRec             & 3.67$\pm$0.10 	& 1.91$\pm$0.08	& 4.93$\pm$0.05 &	2.27$\pm$0.10	& 6.84$\pm$0.24	& 2.65$\pm$0.07 \\
\cdashline{2-8}
& DimeRec                    & \textbf{4.40$\pm$0.05}* 	& 1.92$\pm$0.02	& \textbf{6.48$\pm$0.04}* &	\textbf{2.44$\pm$0.02}*	& \textbf{10.35$\pm$0.19}*	& \textbf{3.20$\pm$0.09}* \\

\midrule
\multirow{8}{*}{KuaiRec}
& SASRec                  & 4.26$\pm$0.11 	& 3.13$\pm$0.13	& 5.62$\pm$0.09 &	3.47$\pm$0.07	& 8.92$\pm$0.15	& 4.12$\pm$0.12 \\
& GRU4Rec                 & 3.27$\pm$0.12 	& 1.85$\pm$0.08	& 4.79$\pm$0.11 &	2.24$\pm$0.04	& 8.73$\pm$0.23	& 3.01$\pm$0.05 \\
& MIND                 & 3.42$\pm$0.06 	& 1.85$\pm$0.05	& 5.23$\pm$0.12 &	2.31$\pm$0.10	& 8.21$\pm$0.19	& 2.89$\pm$0.02 \\
& ComiRec                  & 6.26$\pm$0.07 	& \textbf{4.61$\pm$0.19}	& 7.40$\pm$0.17 &	4.61$\pm$0.11	& 9.98$\pm$0.18	& 5.40$\pm$0.05 \\
& Mult-VAE             & 3.45$\pm$0.05 	& 1.88$\pm$0.05	& 5.67$\pm$0.10 &	2.81$\pm$0.12	& 9.12$\pm$0.17	& 3.98$\pm$0.02 \\
& DreamRec                 & 4.72$\pm$0.05 	& 4.09$\pm$0.04	& 5.23$\pm$0.03 &	4.22$\pm$0.07	& 5.92$\pm$0.06	& 4.35$\pm$0.04 \\
& DiffuRec             & 2.98$\pm$0.03 	& 1.81$\pm$0.07	& 4.39$\pm$0.11 &	2.16$\pm$0.20	& 7.88$\pm$0.09	& 2.85$\pm$0.12 \\
\cdashline{2-8}
& DimeRec                    & \textbf{6.46$\pm$0.05}* 	& 4.71$\pm$0.06	& \textbf{8.23$\pm$0.12}* &	\textbf{5.16$\pm$0.10}*	& \textbf{12.19$\pm$0.13}*	& \textbf{5.93$\pm$0.11}* \\

\midrule
\multirow{8}{*}{ML-10M}
& SASRec                  & 7.72$\pm$0.35 	& 3.58$\pm$0.12	& 14.50$\pm$0.23 &	5.28$\pm$0.17	& 30.02$\pm$0.65	& 8.32$\pm$0.15 \\
& GRU4Rec                 & 10.36$\pm$0.26 	& 4.78$\pm$0.13	& 18.44$\pm$0.31 &	6.81$\pm$0.15	& 35.42$\pm$0.73	& 10.15$\pm$0.25 \\
& MIND                 & 8.42$\pm$0.29 	& 3.88$\pm$0.15	& 15.82$\pm$0.12 &	5.74$\pm$0.19	& 29.70$\pm$0.69	& 8.49$\pm$0.11 \\
& ComiRec                  & 4.36$\pm$0.15 	& 1.68$\pm$0.09	& 12.24$\pm$0.15 &	3.65$\pm$0.11	& 31.22$\pm$0.68	& 7.40$\pm$0.05 \\
& Mult-VAE             & 6.74$\pm$0.13 	& 3.23$\pm$0.08	& 12.56$\pm$0.11 &	4.91$\pm$0.20	& 31.51$\pm$0.59	& 7.99$\pm$0.12 \\
& DreamRec                 & 0.10$\pm$0.02 	& 0.04$\pm$0.01	& 0.26$\pm$0.03 &	0.08$\pm$0.01	& 0.52$\pm$0.06	& 0.13$\pm$0.01 \\
& DiffuRec             & 9.60$\pm$0.31 	& 4.33$\pm$0.15	& 17.00$\pm$0.21 &	6.19$\pm$0.20	& 33.58$\pm$0.94	& 9.46$\pm$0.09 \\
\cdashline{2-8}
& DimeRec                    & \textbf{14.86$\pm$0.22}* 	& \textbf{5.98$\pm$0.16}*	& \textbf{27.58$\pm$0.25}* &	\textbf{9.18$\pm$0.14}*	& \textbf{49.28$\pm$1.27}*	& \textbf{13.49$\pm$0.10}* \\

\bottomrule
\end{tabularx}
\label{tab:main_table}
\end{table*}
\subsubsection{Training and Evaluation Protocal}
We use the historical interaction sequences of users, sorted by timestamp, to predict the next item with which a user might interact. Similar to the setup in most existing works, we partition the users for the training, validation and test sets with a ratio of 8:1:1. We adopt two widely used metrics, HR (Hit Rate) and NDCG (Normalized Discounted Cumulative Gain), to evaluate all methods. HR reflects the model's ability in retrieval, while NDCG indicates the model's ranking ability.

\subsubsection{Baseline Methods}
We compare the proposed DimeRec with three types of sequence recommendation models, including traditional sequence models, multi-interest models, and generative models. The baselines are introduced as follows. \textbf{SASRec}\cite{Kang_McAuley_2018} utilizes a unidirectional Transformer Encoder as its structure and captures user interests through self-attention mechanism. \textbf{GRU4Rec}\cite{hidasi2015session} utilizes recurrent neural network for sequence recommendation. \textbf{MIND}\cite{Li_Liu_Wu_Xu_Zhao_Huang_Kang_Chen_Li_Lee_2019} utilizes capsule network as a multi-interest extractor, decouples user's behavior sequence into diverse interests. \textbf{ComiRec}\cite{Cen_Zhang_Zou_Zhou_Yang_Tang_2020} utilizes the attention mechanism as a multi-interest extractor and allows a controllable diversity retrieval. Our proposed DimeRec also builds upon it. \textbf{Mult-VAE}\cite{liang2018variational} introduce a generative model with a multinomial likelihood function parameterized by neural network. \textbf{DreamRec}\cite{yang2023generate} takes the output of SASRec as the input for the Diffusion Denoising module and eliminates negative sampling, training solely with positive samples. \textbf{DiffuRec}\cite{Li_Sun_Li_2023} attempts to combine generative diffusion models with sequential recommendations by reconstructing the target item embedding with a Transformer-based Approximator.

\subsubsection{Implementation Details}
We implement all of the experiments with Pytorch 1.11 in Python 3.8. We use ComiRec as our Model-based GEM module. The code\footnote{https://anonymous.4open.science/r/DimeRec-154D/} mainly builds upon the \textbf{ComiRec} \footnote{https://github.com/THUDM/ComiRec} codebase. We set same value for basic parameters. The size of embedding $d$ is set to 64, the batch size is set to 256. For each positive sample, we uniformly sample 10 items that the user has not interacted with as negative samples as in the prior works. For all of the multi-interest models, we set the interest number $K$ to 4. We use the Adam optimizer with learning rate 0.005 for all models.

\subsection{Overall Performance}

\subsubsection{\textbf{Recommendation metrics evaluation}}
In this section, we compare the HR and NDCG of DimeRec and all baselines when retrieving 10, 20, and 50 items. All the results are shown in Table \ref{tab:main_table}. The following observations from Table \ref{tab:main_table} answer the \textbf{RQ1}.

\textbf{DimeRec} significantly outperforms all baselines on all datasets w.r.t HR@20, HR@50, NDCG@20 and NDCG@50. Taking the performance of retrieving 50 items as an example, compared to the second-best method, DimeRec's improvement in HR@50 is 18.29\%, 22.14\%, and 39.13\%, and the improvement in NDCG@50 is 20.75\%, 9.81\%, and 32.90\%, on YooChoose, KuaiRec and ML-10M datasets, respectively. DimeRec also shows significant improvement in HR@10, but in terms of NDCG@10, it only achieves comparable results on the YooChoose dataset. This is because the proposed DimeRec is a model for "matching stage", which learns the distribution of a user's next interests in the item space through the diffusion process. As such, it is not as sensitive to ranking metrics, particularly in the ranking of top items. However, this does not affect its effectiveness as a retrieval model, especially when considering that the number of items retrieved in industrial applications is much more than 50.

\textbf{DimeRec} performs much better than existing multi-interest models and Diffusion-based models. We use ComiRec\cite{Cen_Zhang_Zou_Zhou_Yang_Tang_2020} as DimeRec's GEM module, and after introducing the diffusion module, it performs better than the standard ComiRec. This indicates that fitting the distribution of interests in the item space is more effective compared to learning fixed user interests representation. Compared to existing diffusion models, we condition on the user's interests rather than the user's item interaction sequence. Under the guidance of relatively coarse granularity, and more stable interests, we deduce the user's next interest by denoising Gaussian random noise. The result in the table shows that this paradigm is more effective.

\begin{table}[t]
    \centering
    \caption{Linear Probing Accuracy and Average Number of Categories of different models. Note that the Average Number of Categories of DreamRec is crossed out since it struggles to work on ML-10M, with its recommendation performance approaching that of a random strategy. Therefore, even though it exhibits diversity, the recommended items are mostly meaningless.}
    \resizebox{\textwidth}{!}{
    \begin{tabular}{ccccccccc}
    \toprule
        Method &  SASRec & GRU4Rec & MIND & ComiRec & Mult-VAE & DreamRec & DiffuRec & \textbf{DimeRec}\\
    \midrule
        Linear Probing Accuracy & 0.4947 & 0.4743 &0.4699 &0.4917 & 0.4762 &0.2697 & 0.4675 & \textbf{0.5683}\\    \midrule
        Average Number of Categories & 11.40 & 11.23  &11.67 &11.74 & 11.41 & \sout{14.01} & 11.79 & \textbf{11.89}\\
    \bottomrule
    \end{tabular}}
    \label{tab:linear_probing}
\end{table}

\subsubsection{\textbf{Validation for DimeRec's effectiveness}} 
From last section, we know that the diffusion model can enhance the overall performance of recommendation models, especially when comparing \textbf{DimeRec} with ComiRec. In this section, we conduct experiments to explore why DimeRec performs so well to answer \textbf{RQ2}. 

Recent research\cite{chen2024deconstructing} in image generation has found that Diffusion Models not only make progress in generative tasks but also possess strong capabilities in representation learning. Furthermore, DimeRec learns the distribution of a user's next interests through the DAM module, which is more uncertain and diverse than a deterministic interest vector. Based on these two perspectives, We adopt \textit{linear probing} to validate the quality of representation learning, and record the \textit{average number of categories} included in the top 100 recommended items to verify the diversity of recommendations.

Using the item category features that come with the ML-10M dataset, we first train different models, and then we freeze the item embeddings from the converged models. Finally, we add a trainable linear layer to perform classification training on the item embeddings. The classification accuracy obtained from different models is summarized in Table \ref{tab:linear_probing}. We can conclude that \textbf{DimeRec}'s linear probing accuracy of the classification task significantly outperforms other models, which indicates that \textbf{DimeRec} can mine hidden information of items (such as categories) from the interaction history, thereby making the learned item representation space more elegant and ultimately enhancing the recommendation performance. 

From Table \ref{tab:linear_probing}, we can see that DreamRec recommends the most categories since it can not work very well on this dataset, despite our efforts to fine-tune it. Its recommended items are almost equivalent to a random strategy, which is highly diverse. However, the primary goal of a recommendation system is to ensure that the recommended items align with user interests. Thereby, neglecting DreamRec, \textbf{DimeRec} becomes the model which recommends the highest number of categories, which validates that DimeRec can increase the diversity of recommendations.

\subsection{Ablation Study}
\subsubsection{\textbf{Ablation for GEM}}
We replace the GEM module in DimeRec with different models, including the Transformer-based SASRec, a simple MLP, and the dynamic-routing based ComiRec(ComiRec-DR). The self-attention based ComiRec (ComiRec-SA) is the method used in our work. We summarize the results of Recall@20 in Table \ref{tab:gem}. The performance of the first two methods is inferior to that of the latter two demonstrating that the multi-interest model can enhance the learning of the Diffusion model, thereby validating the advantages of generating user's next interest. The results also show that the ComiRec-SA performs better than ComiRec-DR.
\begin{table}[t]
    \centering
    \caption{The performance (Recall@20) of different models as the Guidance Extraction Module in DimeRec.}

    \begin{tabular}{ccccc}
    \toprule
        GEM Module &  SASRec & MLP &  ComiRec-DR & ComiRec-SA\\
    \midrule
        Yoochoose & 0.0587 & 0.0419 & 0.0602 & \textbf{0.0648} \\
        KuaiRec & 0.0688 & 0.0533  &0.0789 & \textbf{0.0823} \\
        ML-10M & 0.2291 & 0.1915 &0.2447 &\textbf{0.2758} \\  
    \bottomrule
    \end{tabular}
    \label{tab:gem}
\end{table}
\subsubsection{\textbf{Ablation for DAM and Loss Function}}
\begin{figure}[htb]
	\centering
 	\includegraphics[width=0.45\textwidth]{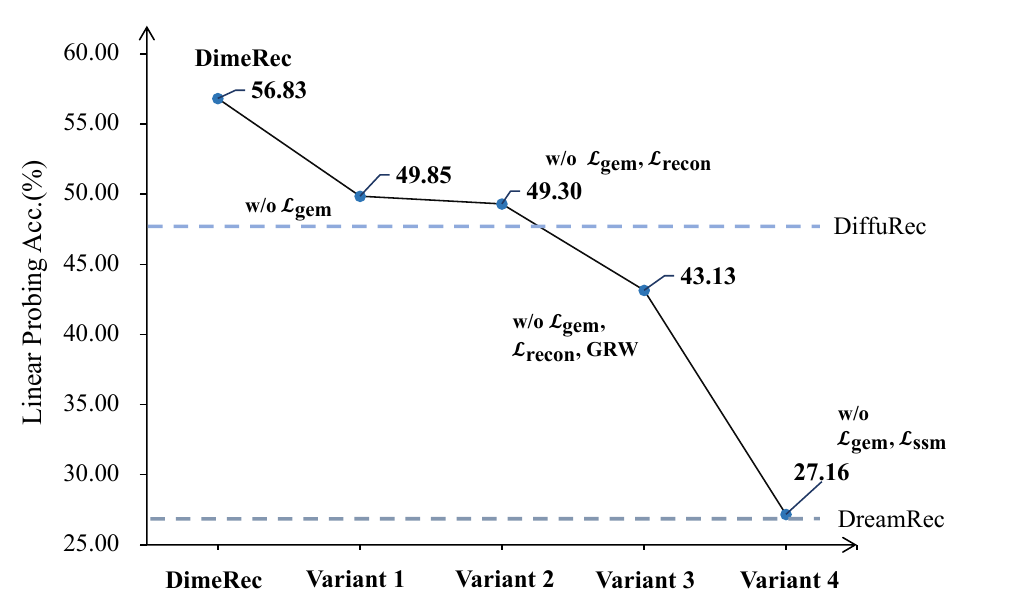}
        \vspace{-1em}
	\caption{Ablation Study: The influence of ablating different components on Linear Probing. The two blue horizontal lines represent the results of DiffuRec and DreamRec, respectively.}
	\label{fig:ablation}
\end{figure}

In this section, we examine the effectiveness of Diffusion module and the new loss function, including the three losses in Equation \ref{eq:total_loss} and Geodesic Random Walk (GRW) in DAM. We perform four types of ablations on DimeRec, specifically: 
\begin{itemize}
    \item Variant 1: DimeRec w/o $\mathcal{L}_{gem}$
    \item Variant 2: DimeRec w/o $\mathcal{L}_{gem}$ and $\mathcal{L}_{recon}$
    \item Variant 3: DimeRec w/o $\mathcal{L}_{gem}$, $\mathcal{L}_{recon}$, and GRW
    \item Variant 4: DimeRec w/o $\mathcal{L}_{gem}$ and $\mathcal{L}_{ssm}$
\end{itemize}
The results of these four ablations, along with the original DimeRec's linear probing results, are summarized in Figure \ref{fig:ablation}. The following conclusions summarized from the figure can answer the \textbf{RQ3}.

By comparing the original DimeRec with Variant 1, we can determine that $\mathcal{L}_{gem}$ significantly impacts DimeRec.  This illustrates the importance of a guidance loss that is independent of the Diffusion module, which is precisely what existing Diffusion-based recommendation models lack. In an end-to-end trained recommendation model, the item representations are randomly initialized,so the contained information is chaotic. In contrast, image generation tasks use either the original pixels of images or a pre-trained encoder for image representation, resulting in relatively stable and meaningful representations. Therefore, the introduction of $\mathcal{L}_{gem}$ is necessary as it allows for more stable learning of item representations.

Comparing Variant 1 and Variant 4, it is evident that removing $\mathcal{L}_{ssm}$ significantly reduces DimeRec's performance. 
Variant 4 shares certain similarities with DreamRec because both are trained only with $\mathcal{L}_{recon}$.$\mathcal{L}_{ssm}$ calculates the score between the final generated user interest embedding and the target item, and it utilizes the sample softmax loss to prioritize the score of positive samples. It can be said that $\mathcal{L}_{ssm}$ controls the final crucial step in DimeRec, underlining its undeniable importance.

\begin{figure}[htb]
	\centering
 	\includegraphics[width=0.35\textwidth]{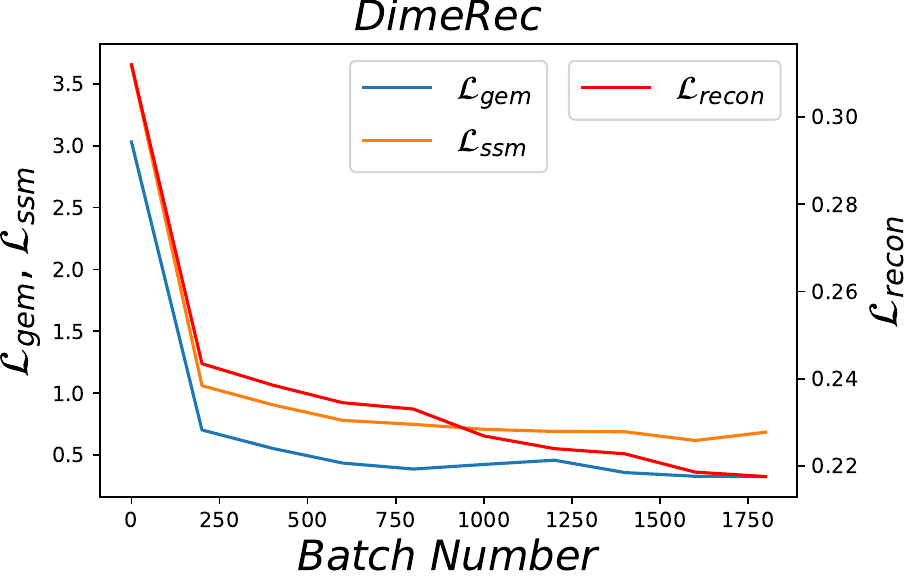}
 	\includegraphics[width=0.35\textwidth]{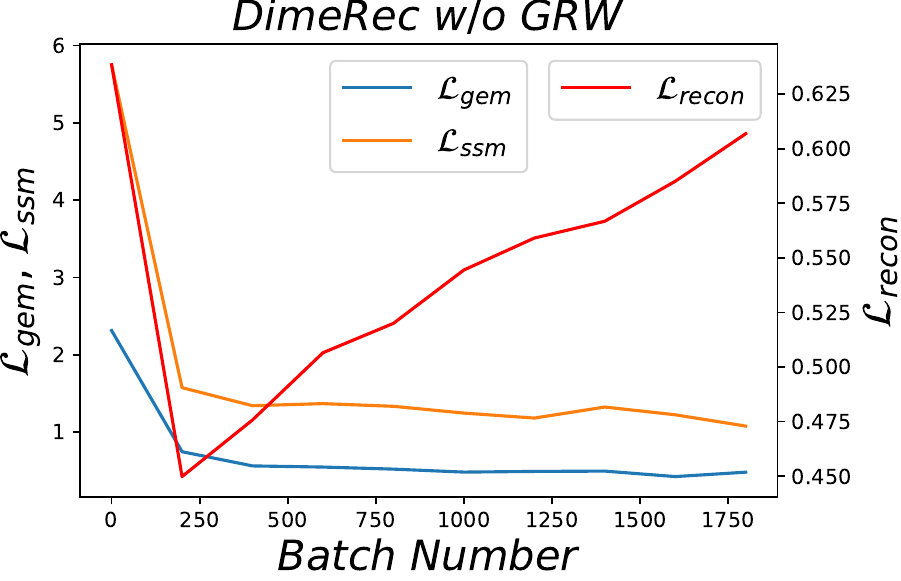}
  \vspace{-0.5em}
	\caption{The influence of GRW on the three losses during the training process. The blue line represents $\mathcal{L}_{gem}$, the orange line represents $\mathcal{L}_{ssm}$, and the red line represents $\mathcal{L}_{recon}$.}
	\label{fig:ab_GRW}
\end{figure}

Comparing Variant 1 with Variant 2 reveals that $\mathcal{L}_{recon}$ also positively impacts DimeRec, though to a lesser extent. However, removing GRW from Variant 2, which becomes Variant 3, reduces the accuracy from 49.85\% to 43.13\%, indicating that constraining item representations to a spherical space and adding noise through Geodesic Random Walks is crucial for Diffusion-based recommendation systems. Based on this observation, we conduct another ablation experiment by removing GRW from the DimeRec and record the changes in the three losses during training. The results are summarized in Figure \ref{fig:ab_GRW}. From the Figure, we can see that during the training of the original DimeRec, all three losses decrease simultaneously as the number of training batches increases. However, upon removing GRW, $\mathcal{L}_{recon}$ gradually increases in the later stages of training. This indicates that in Euclidean space, $\mathcal{L}_{recon}$ is incompatible with $\mathcal{L}_{gem}$ and $\mathcal{L}_{ssm}$, but when we transition to spherical space and apply noise with GRW, this issue can be resolved.
To more clearly and concisely express the function of each component, we also conduct another type of ablation experiment. In this experiment, we independently removed one component at a time. The Recall@10 results on ML-10M are summarized in Table \ref{tab:new_ab}. As shown in the table, the ablation of each individual module negatively impacts the performance, verifying their indispensability.

\begin{table}[t]
    \centering
    \setlength{\tabcolsep}{2.5pt} 
    \caption{The performance of DimeRec after ablating one component. Full means original DimeRec.}
    \begin{tabular}{cccccc}
    \toprule
        Variant &  w/o $\mathcal{L}_{gem}$ & w/o $\mathcal{L}_{recon}$ &  w/o $\mathcal{L}_{ssm}$ & w/o GRW & Full\\
    \midrule
        Recall@20 & 0.0634 & 0.0776 & 0.0630 & 0.0598 & \textbf{0.0823}  \\
        NDCG@20 & 0.0386 & 0.0484  &0.0431 & 0.0370 & \textbf{0.0516} \\
    \bottomrule
    \end{tabular}
    \label{tab:new_ab}
\end{table}

\begin{figure}[htb]
	\centering
 	\includegraphics[width=0.32\textwidth]{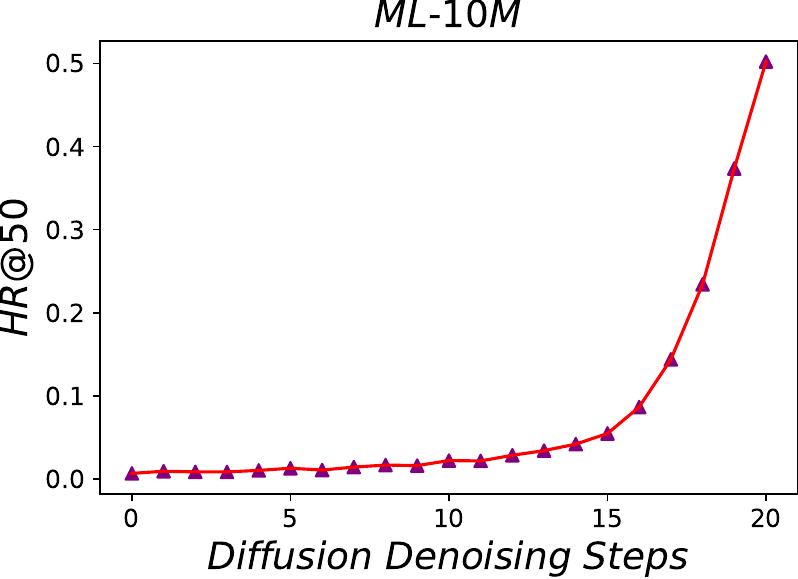}
 	\includegraphics[width=0.32\textwidth]{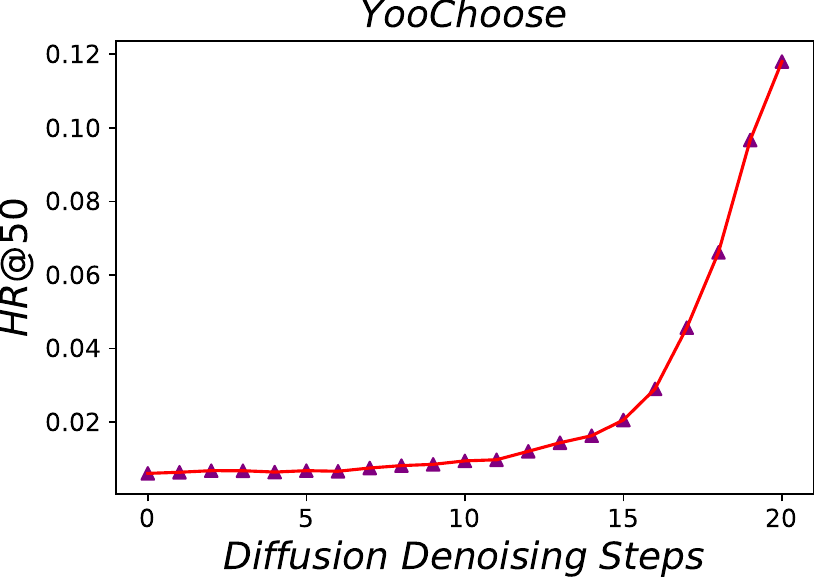}
	\caption{HR@50 calculated by using the intermediate embeddings from the denoising process on ML-10M and YooChoose Dataset. The results on KuaiRec show the same trend.}
	\label{fig:ab_steps}
\end{figure}

Now, we set different denoising steps during the inference phase and use the intermediate embeddings generated by the DAM module for retrieval. The HR@50 obtained is recorded in Figure \ref{fig:ab_steps}. From the figure, it can be seen that with an increase in denoising steps, HR@50 gradually increases, consistent with what we expected. 

\subsection{Hyper-parameter Sensitivity Analysis}
In this section, we investigate the impact of Max Diffusion Step $T$ to answer \textbf{RQ4}. We conduct evaluations on \textbf{DimeRec} by varying the Max Diffusion Step in \{2,10,20,50,100\}. The results are shown in Table \ref{tab:T}. From the results, it can be seen that setting the number of steps to 20 yields the best results, which is one to two orders of magnitude smaller than the step settings in DreamRec, which helps us to deploy online with almost no need to sacrifice efficiency. This is attributed to our transformation of the generation target from item representation to user interest representation, which has a distribution that is easier for the diffusion model to learn. We also conduct the parameter sensitivity analysis on $\lambda$ and $\mu$ in the loss function. Please refer to the appendix for the relevant results.

\begin{table}[htb]
    \centering
    \caption{The HR@20 obtained by DimeRec with different Max Diffusion Steps}
    \begin{tabular}{cccccc}
    \toprule
    \hline
        $T$ &  5 & 10 & 20 & 50 & 100 \\
    \midrule
         YooChoose &0.0412  & 0.0586 &\textbf{0.0648} & 0.0640 & 0.0612   \\
         KuaiRec & 0.0613& 0.0794& \textbf{0.0823} & 0.0775 & 0.0694 \\
         ML-10M & 0.1722 & 0.2341 & \textbf{0.2758} & 0.2529  & 0.2246\\
    \hline
    \bottomrule
    \end{tabular}
    \label{tab:T}
\end{table}

\subsection{Online Results}
We conduct strict online A/B tests in a real-world industrial streaming short-video recommender system to answer \textbf{RQ5}. In the experimental group, we deployed \textbf{DimeRec} as a new source in the retrieval stage, compared with the base group (including ComiRec, Collaborative Filtering, and DNN-based retrieval methods). Both the experimental and base groups are observed on a $10\%$ traffic with about 13 million users, and the observation period was set to 7 days. Table \ref{tab:ab_stat} summarizes the overall metric gains of \textbf{DimeRec} including the average Time Spent over all users and diversity in terms of category counts over effectively watched videos. \textbf{DimeRec} has achieved significant business gain, and been deployed to the main scene traffic, serving over hundreds of millions of active users.

\begin{table}[htb]
    \centering
    \caption{Online A/B results of the experimental group (w/ DimeRec) over base group (w/o DimeRec).}
    \begin{tabular}{c|c|c}
    \toprule
        Metric &  Absolute Improvement & Confidence Interval \\
    \midrule
         Time Spent & +0.110\% & [0.05\%, 0.17\%] \\
         Category Num & +0.139\% & [0.09\%, 0.16\%] \\
    \bottomrule
    \end{tabular}
    \label{tab:ab_stat}
\end{table}
\section{Conclusion}
\label{sec:conclusion}

In this paper, we introduce DimeRec for generating users' next interest of items. It captures the interests from users' historical interaction behaviors through a multi-interest model and then generates the next possible interest through a diffusion model on the spherical representation space under the guidance of the captured interests. Finally, it adopts the generated interest to find the closest item. We have conducted extensive offline experiments on 3 public datasets and online experiments on a short video platform. A series of results confirms the effectiveness of DimeRec, giving us reason to believe that it has become a new SOTA SR model.


\newpage
\printbibliography

\newpage
\appendix
\section{Pseudocode}
\subsection{pseudocode for training}
\begin{algorithm}

\caption{DimeRec Training}
\label{alg:algo_2}
\SetKwComment{Comment}{\# }{}
\KwIn{
\\
\ \ Historical sequence  $\mathcal{S}^u$; Target item embedding $\mathbf{e}_a$; Max diffusion step $T$; Pre-designed diffusion noise schedule $\beta$; Random initialized GEM module $G_{\phi}$; Random initialized DAM module: $D_{\theta}$;
}
\KwOut{
\\
\ \ Well-trained $\textbf{GEM}_{\phi}$ and $\textbf{DAM}_{\theta}$.
}
\SetKwFunction{}
\SetAlgoLined
\BlankLine
\While{ not converged}{
\Comment{GEM}
$\mathcal{H}^u \longleftarrow \mathcal{F}\left(\mathcal{S}^u\right)$ \;
extract guidance through $\mathbf{g}^u \longleftarrow G_{\phi}(\mathcal{H}^u)$ \;
\Comment{DAM}
sample timestep $t \sim \text{Uniform}[0, T]$ \;
Map the embedding onto a sphere: $\mathbf{e}_a' \leftarrow L_2\_Norm(\mathbf{e}_a)$ \;
Sample a Gaussian in $\mathbf{e}_a'$'s tangent space: $\epsilon\sim\mathcal{N}(0, \mathbb{I})$\;
\Comment{$\exp_x[v] = \cos(\|v\|)x + \sin(\|v\|)\frac{v}{\|v\|}$}
Move along the geodesic: $x_t \longleftarrow
\exp_{\mathbf{e}_a'}[\sqrt{1 - \bar{\alpha}_t} \epsilon]$\;

reconstruction $\hat{x}_0 = f_{\theta}(x_t, t, \mathbf{g}^u)$ \;
optimize $\theta, \phi$ with loss $\mathcal{L}$ via Eq. \pref{eq:total_loss} with Adam \;
}
\end{algorithm}

\subsection{pseudocode for serving}
\begin{algorithm}
\caption{DimeRec Reversing}
\label{alg:algo_3}
\SetKwComment{Comment}{\# }{}
\KwIn{
\\
\ \ Historical sequence  $\mathcal{S}^u$; Max diffusion step $T$; Pre-designed diffusion noise schedule $\beta$; Well-trained GEM module $G_{\phi}$; Well-trained DAM module: $D_{\theta}$;
}
\KwOut{
\\
\ \ User's next interest embedding $e_u$.
}
\SetKwFunction{}
\SetAlgoLined
\BlankLine

$t \longleftarrow T$\;
\While{t>0}{
$\mathcal{H}^u \longleftarrow \mathcal{F}\left(\mathcal{S}^u\right)$ \;
extract guidance through $\mathbf{g}^u \longleftarrow G_{\phi}(\mathcal{H}^u)$ \;
reconstruction $\hat{x}_0 = f_{\theta}(x_t, t, \mathbf{g}^u)$ \;
move one step $x_{t-1}\longleftarrow L_2\_Norm(\tilde{\mu}_t(x_t, x_0) + \tilde{\beta}_t \cdot \epsilon'), \epsilon'\sim\mathcal{N}(0,\mathbb{I})$\;
iteration $t\longleftarrow t - 1$\;
}
$e_u\longleftarrow x_0$\;
\end{algorithm}
\section{Hyper-parameter Sensitivity Analysis}

In this section, we investigate the impact of $\lambda$ and $\mu$ in the loss function to answer \textbf{RQ4}. We conduct evaluations on \textbf{DimeRec} by varying the coefficient ($\lambda$) of $\mathcal{L}_{recon}$ in \{0.01, 0.1, 1.0, 10\}. The results are shown in Table \ref{tab:lambda}. We can observe that the When $\lambda=0.1$, the best results are achieved on all three datasets. This suggests that the model is not very sensitive to this parameter's variation across different datasets.

\begin{table}[htb]
    \centering
    \caption{The HR@50 obtained by DimeRec with different coefficients($\lambda$) of $\mathcal{L}_{recon}$, $\mu$ is fixed at 10.}
    \begin{tabular}{c|c|c|c|c}
    \toprule
    \hline
        &  $\lambda=0.01$ & $\lambda=0.1$ & $\lambda=1.0$ & $\lambda=10.0$\\
    \midrule
         YooChoose &0.0988 & \textbf{0.1001} & 0.0951  & 0.0874 \\
         KuaiRec & 0.1059 & \textbf{0.1158} & 0.1017 & 0.0978  \\
         ML-10M & 0.4832 & \textbf{0.4928} & 0.4910 & 0.4810 \\
    \hline
    \bottomrule
    \end{tabular}
    \label{tab:lambda}
\end{table}
We also conduct evaluations on \textbf{DimeRec} by varying the coefficient ($\mu$) of $\mathcal{L}_{ssm}$ in \{0.1, 1.0, 10.0, 100.0\}. The results are shown in Table \ref{tab:mu}. On the YooChoose and KuaiRec datasets, setting $\mu$ to 1.0 yields the best results, while on ML-10M, setting $\mu$ to 10 performed the best. Therefore, we recommend researchers to choose a value between 1 and 10 for $\mu$ when using DimeRec.
\begin{table}[htb]
    \centering
    \caption{The HR@50 obtained by DimeRec with different coefficients($\mu$) of $\mathcal{L}_{ssm}$. $\lambda$ is fixed at 0.1.}
    \begin{tabular}{c|c|c|c|c}
    \toprule
    \hline
        &  $\mu=0.1$ & $\mu=1.0$ & $\mu=10.0$ & $\mu=100.0$\\
    \midrule
         YooChoose & 0.0932 & \textbf{0.1035} & 0.0953 & 0.0921  \\
         KuaiRec & 0.0950& \textbf{0.1219}& 0.1158 & 0.1062 \\
         ML-10M & 0.4570 & 0.4638 & \textbf{0.4928} & 0.4788 \\
    \hline
    \bottomrule
    \end{tabular}
    \label{tab:mu}
\end{table}


\end{document}